# Terahertz Optical Properties and Birefringence in Single Crystal Vanadium doped [100] β-Ga$_2$O$_3$

**Ajinkya Punjal[1,2,†], Shraddha Choudhary[1,*], Maneesha Narayanan[1], Ruta Kulkarni[1], Arumugam Thamizhavel[1], Arnab Bhattacharya[1], Shriganesh Prabhu[1,]**

[1]Department of Condensed Matter Physics and Material Science, Tata Institute of Fundamental Research (TIFR), Mumbai – 400005, India.
[2]Department of Mechanical Engineering, Vishwakarma Institute of Information Technology, Pune – 411048, India

**Corresponding Author:** [†]ajinkya.punjal@tifr.res.in, [*]shraddha.choudhary@tifr.res.in

**Abstract:** We report the Terahertz optical properties of the Vanadium doped [100] β-Ga$_2$O$_3$ using Terahertz Time-Domain Spectroscopy (THz-TDS). The V-doped β-Ga$_2$O$_3$ crystal shows strong birefringence in the 0.2-2.4 THz range. Further, phase retardation by the V-doped β-Ga$_2$O$_3$ has been measured over the whole THz range by Terahertz Time-Domain Polarimetry (THz-TDP). It is observed that the V-doped β-Ga$_2$O$_3$ crystal behaves both as a quarter waveplate (QWP) at 0.38, 1.08, 1.71, 2.28 THz, and a half waveplate (HWP) at 0.74 and 1.94 THz, respectively.

## Introduction:

In recent years β-Ga$_2$O$_3$ has been extensively studied due to its material properties such as wideband gap (4.9 eV), high electric field breakdown (8 MV/cm), and good thermal and chemical stability [1],[2]. These properties make β-Ga$_2$O$_3$ a promising candidate for many semiconductor devices like Schottky barrier diodes [3], UV photodetectors [4], field-effect transistors [5], etc. In addition to these applications, the high-frequency applications of β-Ga$_2$O$_3$ are also under active investigation, and a proper understanding of material properties at high frequencies is required. Over the last two decades, advances in the generation and detection of Terahertz (THz) radiation enabled researchers to investigate a wide range of material properties in this wavelength range[6]. In particular, Terahertz time-domain spectroscopy (THz-TDS) has evolved into an effective tool for characterizing the optical properties of materials at THz frequencies [7]. It is a non-destructive, contactless technique suitable even for examining highly fragile materials like thin films [8], biological tissues [9], etc. Many semiconductors such as ZnO [10], GaN [11], Graphene [12] have been characterized extensively using THz-TDS. However, there are only a few THz studies on doped and undoped β-Ga$_2$O$_3$ [13]–[16].

In this work, we used the THz-TDS to report the optical properties of Vanadium doped β-Ga$_2$O$_3$ at THz frequencies. Furthermore, we employed the THz polarimetry (TDP) technique to determine the phase retardation and polarization of the THz wave transmitted through the V-doped β-Ga$_2$O$_3$ crystal. We show that the V-doped β-Ga$_2$O$_3$ behaves as a QWP and an HWP at different THz frequencies. To the best of our knowledge, the Terahertz optical properties of V-doped β-Ga$_2$O$_3$ crystal have never been investigated before.

## Results and Discussion:

Single Crystals of [100] oriented Vanadium doped β-Ga$_2$O$_3$ were grown at TIFR using the optical float zone technique [17]. β-Ga$_2$O$_3$ has a base-centered monoclinic structure with C2/m space group symmetry [18] and the corresponding a, b, c are 12.21Å, 3.03 Å, 5.79 Å respectively and *β*=103.83°.

By carefully tuning the growth parameters, electrically resistive V-doped β-Ga$_2$O$_3$ single crystals can be grown. These V-doped crystals have significantly different optical properties compared to the undoped crystal ones, details of which are given in the supplementary information. A cleaved slice of the V-doped β-Ga$_2$O$_3$ crystal was used in this study, with a thickness of ~0.822 mm. The refractive index of the crystal is in general a complex number and the permittivity of the crystal can be calculated using well-known relations [7]. The permittivity of a crystal is a direct measure of the polarizability of the constituent atomic lattice of the crystal and the electron charge distribution in it under an externally applied electric field. The frequency of this field can be from DC to THz and beyond. The low-symmetry (C2/m) of monoclinic β-Ga$_2$O$_3$ results in strongly anisotropic dielectric permittivity parameters along with different crystal directions. Thus, it is imperative to study the response of this crystal under different orientations with respect to the incident THz polarization direction. We have extensively studied the effect of polarizability in changing the polarization state of the transmitted THz signal through the crystal.

A THz-TDS setup (shown in Figure 3 of the supplementary information) is used for measuring the THz transmission and state of polarization through the V-doped crystal. The setup consists of a femtosecond (fs) pulsed laser (FemtoSource Synergy, 800 nm, 80 MHz, 10 fs) and a standard four-parabolic mirror configuration. The laser beam is divided into a pump and probe beam. The pump beam generates THz radiation using an LT-GaAs-based photoconductive antenna (BATOP GmbH). The probe beam is optically delayed and is used for detection purpose. A 2mm thick ZnTe crystal with <110> orientation is used for detecting the generated THz using the standard electro-optic technique. The entire setup is purged with nitrogen to eliminate any unwanted absorption of the THz due to the water vapor present in the ambient air.

The V-doped crystal was oriented with the b-axis parallel to the incident electric field, and the transmitted THz field was measured. After that, the V-doped crystal was rotated by 90º in the sample plane and the transmitted THz field was measured. Figure 1 (a, b) illustrates the transmitted THz electric fields. Figure 1 (a) shows that the transmitted temporal THz field along θ=0º is shifted by 0.62 ps compared to the one along θ=90º, suggesting a strong birefringence in the V-doped Ga$_2$O$_3$ crystal. Furthermore, Figure 1 (b) illustrates the frequency domain THz spectrum.

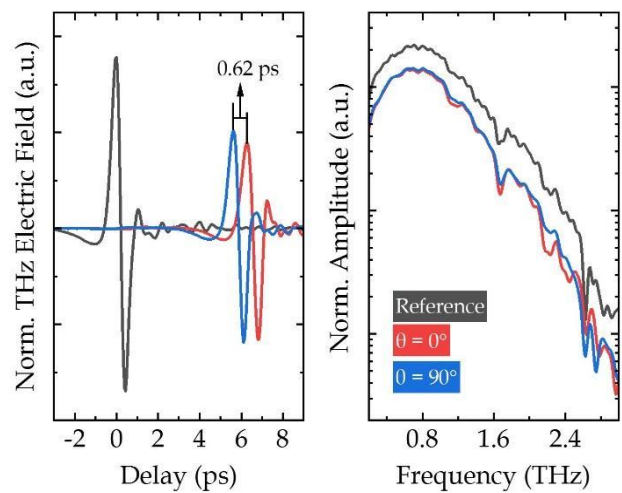

*Figure 1: (a) The transmitted time-domain THz electric field along b-axis of V-doped crystal for θ=0º and θ=90º, (b) Corresponding frequency domain spectra for θ=0º and θ=90º.*

The refractive index and extinction coefficient are calculated to get a better insight into the birefringence of the V-doped crystal. The THz time-domain signals are Fourier transformed to obtain the corresponding frequency domain spectra. The complex-valued transfer function is calculated by normalizing the sample spectrum with the reference spectrum which is then used to determine the complex refractive index (detailed derivation is provided in the supplementary information). Figure 2 (a, b) shows the refractive indices and extinction coefficients for θ=0° and θ=90° with respect to the frequency. Large birefringence is observed across the 0.2 to 2.4 THz range. The refractive index along θ=0° and θ=90° at 1 THz is 3.40±0.01 and 3.10±0.01, respectively. The overall contrast in the refractive index across the frequency range is evaluated to be |0.3±0.02|. To the best of our knowledge, this is the highest among any doped or undoped $Ga_2O_3$ crystals [13], [15]. Furthermore, the extinction coefficient is extremely low (order of $10^{-2}$) throughout the frequency range, indicating that the V-doped crystal has a minimal loss. The high transmittance and minimal loss make such V-doped $\beta$-$Ga_2O_3$ crystals promising candidates for further research into THz frequency applications, especially in the design of polarizing optical elements.

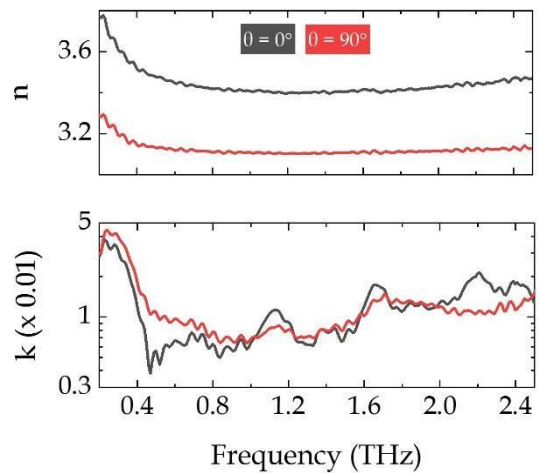

*Figure 2: (a) The calculated complex refractive index for θ=0° and θ=90°, (b) The corresponding extinction coefficient for θ=0° and θ=90°.*

To map the polarisation of the THz field modified by the V-doped crystal, a THz-Time Domain Polarimetry approach similar to [19] is employed. Figure 3 (a) shows a simplified schematic of the setup. The incident THz field is parallel to the x-axis. The V-doped crystal is mounted on a motorized rotating stage and is oriented with the b-axis being parallel to the electric field. The V-doped crystal is followed by a pair of linear THz wire grid polarisers (Tydex) in Polarizer-Analyser configuration. The Polarizer is oriented at an angle ϕ with respect to the x-axis and is mounted on a motorized rotation stage. The Analyser is fixed and positioned along the x-axis. This technique allows us to precisely determine the phase retardation induced by the V-doped crystal without rotating the ZnTe. The V-doped crystal is rotated by an angle θ, and for each θ, the Polarizer (ϕ) is rotated by complete 360 degrees in 10° steps. When the incident THz field is transmitted through the V-doped crystal, the electric field can be expressed as

$$\vec{E} = a\,cos(\omega t)\hat{x} + b\,cos(\omega t + \delta)\,\hat{y} \qquad (1)$$

Where a, b, and $\delta$ are co-polarisation, cross-polarisation, and phase retardation, respectively. As shown in Figure 3 (b), the THz field after passing through the V-doped crystal is successively projected along with the Polarizer and Analyzer [20]. The resulting electric field on the ZnTe detector is given as,

$$S(\phi) = |Cos\phi|\,[(aCos\phi + bSin\phi Cos\delta)^2 + (bSin\phi Sin\delta)^2]^{\frac{1}{2}} \qquad (2)$$

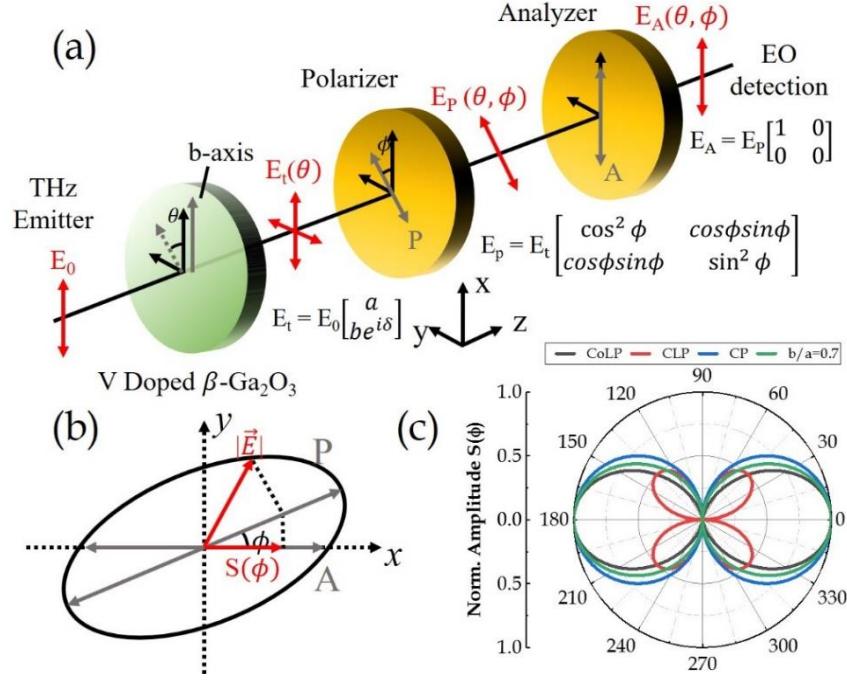

*Figure 3 (a) THz TDP setup schematic. (b) Electric field E projections leading to the measured signal S(ϕ). (c) Simulated amplitude S(ϕ), for the co-and cross-linear polarization (CoLP, CLP), circular polarization (CP) and elliptical polarization (b/a).*

The polarization parameters are obtained by measuring $S(\phi)$ as a function of the polarizer angle $\phi$. If the crystal acts as a QWP, then a=b, $\delta = \pi/2$, and the equation (2) will be simplified to $S(\phi) = a|Cos\phi|$. However, if the crystal acts as an HWP then for an x-polarised incident electric field, we will have a y-polarised electric field at the output; thus a=0, $\delta = \pi$, and the equation (2) will be simplified to $S(\phi) = b/2|Sin2\phi|$. Furthermore, we will have elliptically polarized light for a ≠ b and $\delta = \pi/2$. All of the preceding cases are demonstrated in Figure 3 (c).

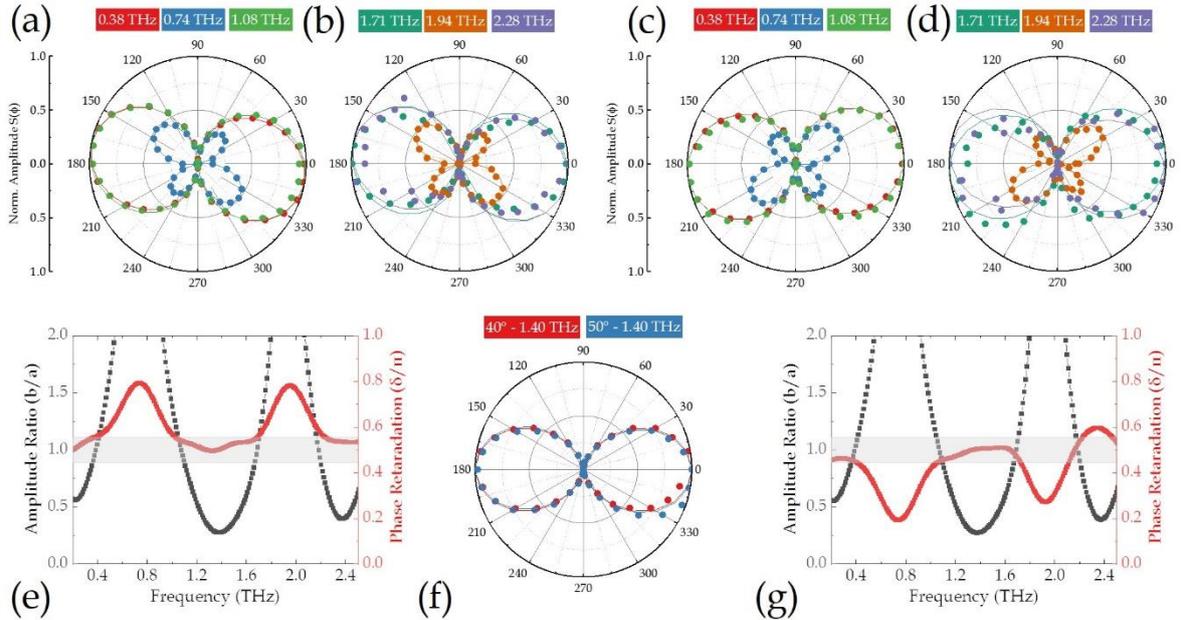

*Figure 4: Polar map for normalized amplitude S(ϕ) at various THz frequencies. (a, b) 0.38, 0.74, 1.08, 1.71, 1.94, and 2.28 THz for θ = 40º. (c, d) 0.38, 0.74, 1.08, 1.71, 1.94, and 2.28 THz for θ = 50º. (e, g) As a function of frequency, the corresponding fitted polarization parameters are plotted. The black curve represents the amplitude ratio b/a, while the red curve illustrates phase retardation δ/π. (f) Polar map at 1.40 THz for θ = 40º and θ = 50º.*

For each value of $\phi$, a full scan over 360° rotation of the Polarizer is performed (in 10° steps), and the corresponding THz field is measured. Polarization parameters of the THz field transmitted from the V-doped $Ga_2O_3$ crystal are obtained by fitting the measured data to equation (2). The normalized polar plots for the measured data and corresponding fitting are shown in Figure 4. As seen from Figure 4 (a, b), the V-doped crystal behaves as a QWP at 0.38, 1.08, 1.71, and 2.28 THz and as an HWP at 0.74, 1.94 THz for θ = 40° (see Figure 3 (c) for reference). There is a slight mismatch between the fit and measured data at 1.71 and 2.28 THz due to the presence of water lines at these frequencies. Furthermore, polar plots for θ = 50° are plotted as illustrated in Figure 4 (c, d), which shows a similar trend at the corresponding frequencies. The transition from left to right-handedness is observed in the transmitted THz wave for the linear polarization case. Correspondingly, a slight right tilt in the circular polarization case is also seen for θ = 50° with respect to θ = 40°. Figure 4 (e, g) illustrates the amplitude ratio and phase retardation profile at θ = 40° and θ = 50°, respectively. The limit for both the phase tolerance and amplitude ratio (b/a) lies within ± 10%. The amplitude ratio and phase retardation values are within the tolerance limit at the frequencies where the V-doped crystal behaves as a QWP and HWP, as shown in Figure 4 (e, g). In addition to this, a polar plot at 1.4 THz where the elliptical polarisation behaviour dominates is given at both θ = 40° and θ = 50° in Figure 4 (f).

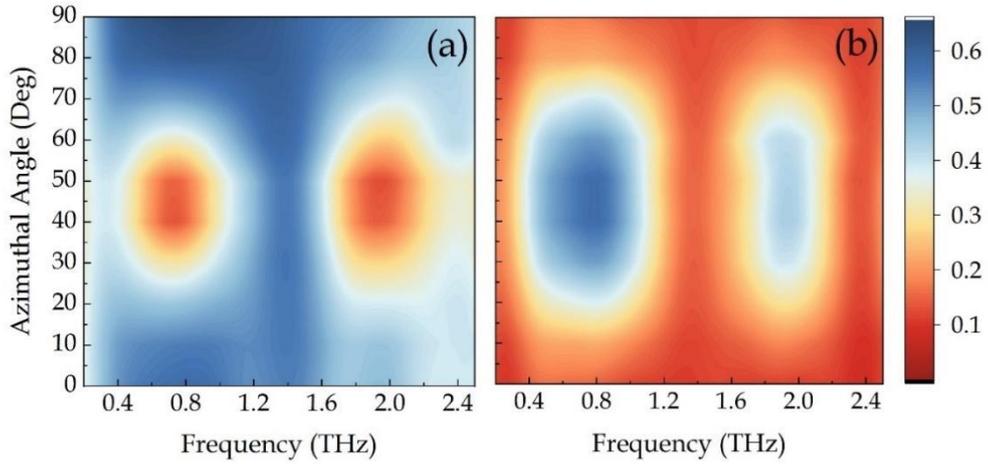

*Figure 5: (a, b) Experimental co- and cross-polarization parameter as a function of azimuthal angle with respect to the frequency, respectively.*

Figure 5 (a, b) illustrates the calculated co- and cross-polarization parameters as a function of azimuthal angle with respect to frequency. Figure 5 shows that the co- and cross-polarization parameters follow a complementary pattern across the frequency range, i.e., when co-polarization is maximum, cross-polarization is minimum, and vice versa. The crystal behaves as an HWP at 0.74 and 1.94 THz, where the co-polarization is minimum and cross-polarization is maximum. Moreover, the QWP type behaviour is seen at 0.38, 1.08, 1.71, and 2.28 THz when the co and cross-polarization are equal, and the phase retardation is $\delta = \pi/2$ as seen previously. For further investigating the polarization state of the transmitted THz wave, we calculated the polarization orientation angle ψ and ellipticity angle χ using the Stokes parameters.

$$S_o^2 = S_1^2 + S_2^2 + S_3^2, \quad (3)$$

Where

$$S_o = a^2 + b^2, \quad (4)$$
$$S_1 = a^2 - b^2, \quad (5)$$
$$S_2 = 2ab\cos\delta, \quad (6)$$
$$S_3 = 2ab\sin\delta. \quad (7)$$

And

$$\Psi = \frac{1}{2}tan^{-1}\left[\frac{S_2}{S_1}\right], \qquad 0 \leq \Psi \leq \pi, \qquad (8)$$

$$\chi = \frac{1}{2}sin^{-1}\left[\frac{S_3}{S_0}\right], \qquad -\frac{\pi}{4} \leq \chi \leq \frac{\pi}{4}. \qquad (9)$$

Figure 6 (a) shows the ellipticity ($\chi$) as a function of frequency, as the azimuth angle is varied over 90°. For certain azimuth angles (θ=45°), the V-doped crystal acts as a QWP at 0.38, 1.08, 1.71, and 2.28 THz. Furthermore, in between these frequencies, a linear polarization state is also observed at 0.74 and 1.94 THz, where it behaves as a half-wave plate. Figure 6 (b) shows the evolution of the orientation angle ($\Psi$) as a function of frequency, as the azimuth angle is varied over 90°. Here, the transition from left-hand polarization to right-hand polarization of the transmitted THz field is clearly visible, which is consistent with the previous findings (see Figure 4 (a, b, c and d)) [21].

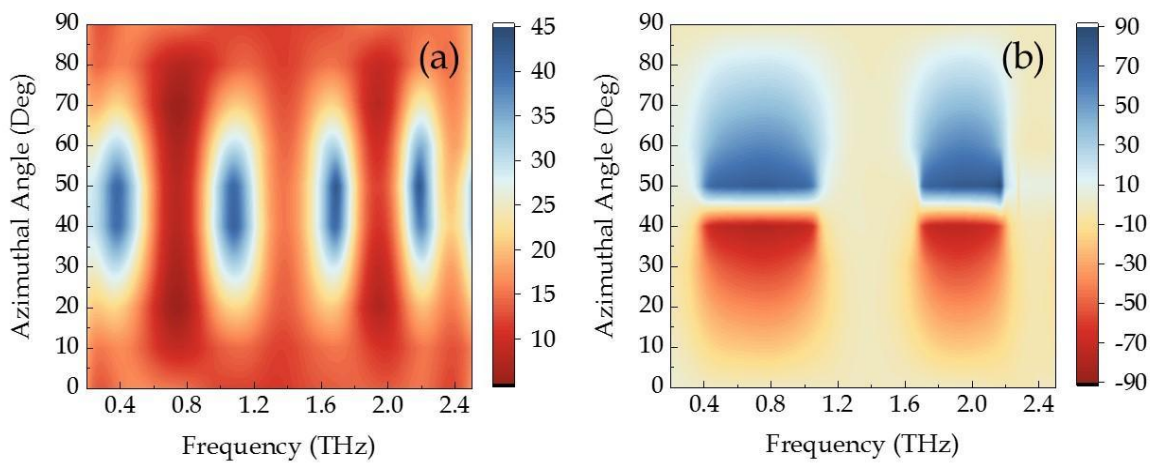

*Figure 6: (a) Ellipticity ($\chi$) as a function of azimuthal angle with respect to the frequency. (b) Orientation angle ($\Psi$) as a function of azimuthal angle with respect to the frequency.*

**Conclusion:**

In conclusion, we report the first Terahertz optical study on V-doped β-$Ga_2O_3$ using the THz-TDS technique. It was observed that V-doped β-$Ga_2O_3$ has significant THz transmission and shows a strong birefringence in the 0.2-2.4 THz range. In addition, a THz polarimetry technique was used to accurately determine the phase retardation and polarization of the transmitted THz field through the V-doped β-$Ga_2O_3$ crystal. The V-doped β-$Ga_2O_3$ crystal behaves as a QWP at 0.38, 1.08, 1.71, and 2.28 THz, while an HWP type behaviour is observed between these frequencies at 0.74 and 1.94 THz, respectively. Due to its high transmission and low losses, V-doped β-$Ga_2O_3$ opens up new possibilities for developing THz QWP, HWP, segmented waveplates, etc., and new opportunities for device integration requiring intrinsic polarisation control of THz light.

**Supplementary Materials:**
See the supplementary information for detailed analysis.

**Data Availability:**
Analysis program of the THz-TDP will be available upon request to the Author (A.P.)

**Authors' Contribution:**
(A.P.) and (S.C.) contributed equally to the THz measurement and data analysis. (A.P.) build and characterized the THz-TDP setup. (M.N.) and (R.K.) has grown the crystals. (A.T.) and (A.B.) supervised the growth of crystals. (S.P.) supervised the overall THz experiments, analysis, and setup build-up.

**Acknowledgment:**
We thank the Department of Atomic Energy (DAE) and Tata Institute of Fundamental Research (TIFR), Mumbai, for the resources, and the work is supported by DAE Grant ID RTI4003.

**Conflicts of Interest:**
The authors declare no conflict of interest.